\documentclass[twocolumn,showpacs,amsmath,amssymb,prl]{revtex4}

\usepackage{graphicx}

\begin{document}

\title{Single-electron transport through the vortex core levels
in clean superconductors}
\author{N. B. Kopnin $^{(1,2)}$}
\author{A. S. Mel'nikov $^{(3,4)}$}
\author{V. M. Vinokur $^{(4)}$}
\affiliation{$^{(1)}$ Low Temperature Laboratory, Helsinki
University of
Technology, P.O. Box 2200, FIN-02015 HUT, Finland,\\
$^{(2)}$ L. D. Landau Institute for Theoretical Physics, 117940
Moscow, Russia\\
 $^{(3)}$ Institute for Physics of Microstructures,
        Russian Academy of Sciences, 603950, Nizhny Novgorod, GSP-105,
 Russia,\\
$^{(4)}$ Argonne National Laboratory, Argonne, Illinois 60439 }

\date{\today}

\begin{abstract}
We develop a microscopic theory of single-electron transport in
N-S-N hybrid structures in the presence of applied magnetic field
introducing vortex lines in a superconductor layer. We show that
vortex cores in a thick and clean superconducting layer are
similar to mesoscopic conducting channels where the bound core
states play the role of transverse modes. The transport through
not very thick layers is governed by another mechanism, namely by
resonance tunneling via vortex core levels. We apply our method to
calculation of the thermal conductance along the magnetic field.
\end{abstract}
\maketitle

Electron transport through various hybrid structures is in the
focus of current nanoscale physics research.  Of a special
interest are normal metal (N) - superconductor (S) - normal metal
(N) trilayers where a superconducting gap $\Delta _0$ suppresses
single-particle transport, making charge transfer transparency
very sensitive to the external controlling parameters. The
electron transmission through such N-S-N structure at low energies
is associated with two-particle Andreev processes.  If the
thickness $d$ of the superconducting slab is much larger than the
coherence length $\xi$ the electrons incident on the slab are
reflected as holes, and the normal current converts into the
supercurrent. Single-electron tunneling through an N-S-N structure
decays exponentially with the slab thickness giving rise, in
particular, to the exponential drop off of the electronic
contribution to the thermal conductance \cite{shikin,demers}. A
single-particle transport through NSN recovers by applying a
strong magnetic field, which creates vortex lines where the gap in
the spectrum is suppressed. In the present Letter we develop a
regular theoretical description of the low-energy single-particle
transfer through the vortex core states in clean type II
superconductors.  Since the single particle contribution to
electric conductivity is short-circuited by supercurrent, we focus
on the thermal conductivity which in this case is the
experimentally accessible characteristic of the one-electron
transport.

Taking the simplest view of a vortex core as a normal channel with
a density of states as in the normal metal, we arrive at a
single-electron Sharvin conductance of a normal wire with the
radius of the order of $\xi$:
  \begin{equation}
     G_{se}\propto (e^2/\pi \hbar)(k_F\xi )^2  \label{Sharvin-cond}
  \end{equation}
where $k_F=p_F/\hbar$ is the Fermi wave vector. A little bit more
attentive second thought shows that in clean superconductors the
only trajectories that do not hit vortex core ``walls'' contribute
to a single-particle conductivity.  Indeed, if a vortex core is a
normal cylinder surrounded by a superconductor, an electron flying
into the core boundary is Andreev reflected as a hole back along
its incidental path. Thus the single-electron transport along such
a trajectory is blocked, and the only contribution to the
conductance of the ``Andreev wire'' comes from trajectories that
traverse freely the normal region. Within this model, such
trajectories must have incident angles $\theta \lesssim \xi /d$,
i.e., are confined to a solid angle $\xi ^2/d^2$. This would
result \cite{MelVin1} in $G_{se}\propto (e^2/\pi
\hbar)(k_F\xi)^2(\xi /d)^2$.  Considering below the realistic gap
profile inside the core in the framework of the developed approach
we will show that the drop of conductance with the thickness $d$
is even more rapid and proportional to $d^{-6}$.

However, the low energy transport associated with the
Caroli--deGennes--Matricon (CdGM) states \cite{degen} propagating
along the core should obviously saturate for very thick
superconducting slabs. The CdGM spectrum $\epsilon_{\mu }(k_z)$ as
a function of the quantized angular momentum $\mu =n+1/2$ varies
from $-\Delta_0$ to $+\Delta_0$ crossing zero as the impact
parameter $b=-\mu/k_{r}$ of the particle varies from $-\infty$ to
$+\infty$. Here $k_r=\sqrt{k_F^2-k_z^2}$ is the wave vector in the
plane perpendicular to the vortex, $(r,\, \phi ,\, z)$ is a
cylindrical coordinate system with the $z$- axis chosen along the
vortex line. For small $\mu$ the spectrum is  $\epsilon_{\mu
}(k_z) \simeq -\mu \Delta_0/(k_r\xi)$. Transport carried by the
quantized transverse modes is described by the Landauer formula.
In the limit $\Delta_0/(k_F\xi)\ll T\ll \Delta_0$, the number of
modes is $\sim (T/\Delta_0 )k_F\xi$, and for the resonant
contribution to the single-particle conductance one gets
\cite{MelVin1}:
\begin{equation}
G_{se}=(e^2/\pi \hbar)\sum _\mu{\cal T}_\mu \sim (e^2/\pi \hbar)
(T/\Delta_0) (k_F\xi ) \ .   \label{Landauer-cond}
\end{equation}
Here $\mu$ numerates the transverse modes with transparencies
${\cal T}_\mu$ open in the vortex core. This estimate can also be
obtained from Eq.\ (\ref{Sharvin-cond}) provided the group
velocity is taken as $v_g =\hbar ^{-1}\partial \epsilon _{\mu}
/\partial k_z\sim \epsilon _{\mu }/\hbar k_F$ instead of the
velocity $v_z \sim v_F$ as in a normal tube. The conductance of
Eq.\ (\ref{Landauer-cond}) is by a factor $(k_F\xi )^{-1}\ll 1$
smaller than that of Eq.\ (\ref{Sharvin-cond}). One thus expects
that the single-particle Andreev-wire conductance of the vortex
core would transform into Eq.\ (\ref{Landauer-cond}) with
increasing $d$.

As we already mentioned, the single-electron transport determines
the behavior of thermal conductivity. The Wiedemann--Franz law
would result in $\kappa \propto TG_{se}/e^2$ for the thermal
conductance along the vortex line. In the vicinity of $H_{c2}$,
the thermal conductivity has been studied theoretically in a
number of papers (see, for instance, \cite{Caroli,Maki,Tes}). In
dirty superconductors, the electron contribution to the thermal
conductance along the vortices for a small concentration of vortex
lines \cite{Kopnin/heat} agrees conceptually with Eq.\
(\ref{Sharvin-cond}): $\kappa (B)\simeq (B/H_{c2}) \kappa_N$,
where $\kappa_N$ is the electron thermal conductivity in the
normal state. Unfortunately, in clean superconductors this simple
estimate fails to describe the experimental data
\cite{Lowell,vinen}: the thermal conductance in the magnetic field
direction appears to be two orders of magnitude smaller. It was
noted first in \cite{vinen} that this obvious conflict can be
caused by a very small group velocity of the CdGM modes as
discussed above.  Analysis of quantum transport through individual
vortices is of particular importance for understanding the
properties of mesoscopic superconductors. The exotic vortex states
in these systems are nowadays the focus of a considerable
attention \cite{geim,Peeters,chibotaru}.

Hereafter we concentrate on the low-energy single-particle
transport through vortex cores in clean ($\ell \gg d$) type II
superconductors in the low-field limit of separated vortices and
develop a systematic approach for calculation of the thermal
conductance along vortex cores. We study the transmission of an
electron wave incident on the superconducting slab placed between
two bulk normal metal electrodes assuming ideally transparent
boundaries and neglecting the normal scattering. Considering two
extremes of infinite and finite slab thicknesses we confirm the
intuitive picture discussed above: For a not very thick slab, the
transmission is determined by the semi-classical resonant
tunneling through the energy gapped region when the energy
coincides exactly with one of the levels in the vortex core. The
transmission is proportional to the large Sharvin conductance;
however, it decays exponentially with the slab thickness except
for the trajectories that go almost parallel to the vortex axis.
The exponential decay is thus replaced with a $(\xi /d)^6$ power
law, which gradually transforms the Andreev-like thermal
conductance into a thickness-independent expression Eq.\
(\ref{Landauer-cond}) as $d$ increases.

{\it Wave functions.}--
Quasiparticles in the superconductor obey the Bogolubov-de Gennes
(BdG) equations:
\begin{eqnarray}
\left[\frac{1}{2m}\left(-i\hbar \nabla -\frac{e}{c}{\bf
A}\right)^2
-E_F\right] u+\Delta v&=&\epsilon u \ , \label{Beq1}\\
\left[\frac{1}{2m}\left(-i\hbar \nabla +\frac{e}{ c}{\bf
A}\right)^2 -E_F\right] v-\Delta ^* u&=&-\epsilon v \ .
\label{Beq2}
\end{eqnarray}
The wave vector $k_z$ along the vortex axis is a good quantum
number, $u=e^{ik_z z}u_{k_z}, \; v=e^{ik_z z}v_{k_z}$. We put
$\Delta =|\Delta |e^{i\phi} $, and $ u_{k_z}=e^{i\phi /2 +i\mu
\phi }U$, $ v_{k_z}=e^{-i\phi /2 +i\mu \phi }V$. Equations
(\ref{Beq1}), (\ref{Beq2}) have eigenvalues $\epsilon _{\mu
}(k_z)$ for the CdGM bound states. We establish first a simple
identity for the localized states. Calculating the derivative with
respect to $k_z$ from the both sides of Eqs.\ (\ref{Beq1}),
(\ref{Beq2}) and using the normalization of the CdGM wave
functions we find
\begin{eqnarray}
&&\int \left[ u^*_{\mu k_z}\left(\hbar
k_z-\frac{e}{c}A_z\right)u_{\mu
k_z} \right. \nonumber \\
&-&\left.  v^*_{\mu k_z}\left(\hbar
k_z+\frac{e}{c}A_z\right)v_{\mu k_z}\right]\, d^2r
=\frac{m}{\hbar}\frac{\partial \epsilon _{ \mu }}{\partial k_z} \
. \label{ident1}
\end{eqnarray}
There is a considerable cancellation in Eq.\ (\ref{ident1}): the
r.h.s. is by a factor of $(k_F\xi )^{-1}$ smaller than each term
in the l.h.s. This is why Eq.\ (\ref{Landauer-cond}) gives much
smaller conductance than Eq.\ (\ref{Sharvin-cond}). We will use
Eq.\ (\ref{ident1}) later to derive the thermal conductance for a
thick slab.

However, the cancellation does not take place for a
finite-thickness slab where the CdGM states are not truly
localized. To consider this in more detail we use a semi-classical
approach and look for $\hat {\cal U} =(U,\, V)$ in the form
\[
\hat{\cal U}= H^{(1)}_l \left(k_r r\right)\hat w^{(+)}
+H^{(2)}_l\left(k_r r\right)\hat w^{(-)}
\]
where $H_l^{(1,2)}$ are the Hankel functions, $l=\sqrt{\mu
^2+1/4}$, assuming that $\hat w=\left(w_{1},\, w_{2}\right)$ are
slow functions of $r$ and $z$. Following \cite{Bardeen} we put $ x
=\sqrt{r^2 -b^2}$, $b=-\mu /k_r $ and define the trajectories $dx
=\pm ds_\pm \sin \theta $, $dz =ds_\pm \cos \theta $ for $\hat
w^{(\pm )}$, respectively, where $k_z =k_F\cos \theta$ and $ds_\pm
$ is the distance along the corresponding trajectory. For a point
$(x,z)$ on the trajectory $z=z_0 \pm x \cot \theta$ we obtain
\begin{eqnarray}
\mp \frac{i\hbar ^2k_r}{m} \frac{d \hat w^{(\pm )}}{d x}
-\hat\sigma _z \left(\epsilon +\frac{\hbar ^2k_r b}{2m(x^2
+b^2)}\right)\hat w^{(\pm )}\nonumber \\
+i\hat\sigma _y|\Delta |\hat w^{(\pm )} =0 \ .\label{eq-w}
\end{eqnarray}
For the functions $w_{1,2}^{(+)}(x,z)$, the limit $x\rightarrow
\infty$ corresponds to a wave radiating from the vortex into the
bulk while $w_{1,2}^{(-)}(x,z)$ corresponds to a wave incident on
the vortex. The condition of regularity at $r=0$  requires $\hat
w^{(+)}(0,z)=\hat w^{(-)}(0,z)$ at the classical turning point,
$x=0$. Let us put $ \hat w^{(+)}(x,z) =\hat w(x,z)$, $\hat
w^{(-)}(x,z) =\hat w(-x,z) $. The functions $w_{1,2}(x,z)$ satisfy
Eq.\ (\ref{eq-w}) with the upper sign along the entire $x$ axis.

\begin{figure}[t]
\centerline{\includegraphics[width=0.7\linewidth]{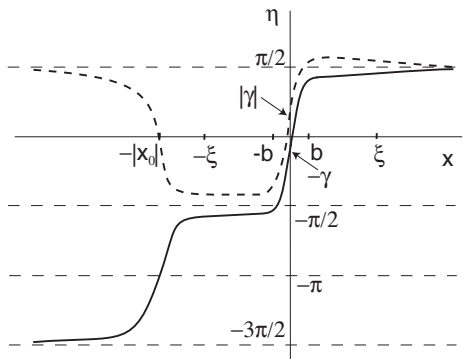}}
\caption{The coordinate dependence of $\eta$ for $\gamma +\eta
_0=0$. The full line is for $\gamma >0$ while the dashed line is
for $\gamma <0$.} \label{fig-eta}
\end{figure}

We introduce new functions $\eta$ and $\zeta$ through $w_1=
e^{\zeta +i\eta /2} $, $w_2= e^{\zeta-i\eta /2} $ and arrive at
the equations
\begin{eqnarray}
\frac{d\eta}{dx} &=&\frac{2m\epsilon}{\hbar ^2k_r}
+\frac{b}{(x^2+b^2)}
-\frac{2m|\Delta |}{\hbar ^2 k_r}\cos \eta \ ,\label{eq-eta2} \\
\frac{d\zeta}{dx} &=&-m|\Delta |\hbar ^{-2}k _r^{-1}\sin \eta \ .
\label{eq-zeta2}
\end{eqnarray}
The requirement that $w$ vanishes at $x\rightarrow \pm \infty$ is
$ \eta =\pm \pi /2- (\epsilon /|\Delta _0) +2\pi k $. These
values, however, are not stable for a general choice of the
integration constants. A general solution for not very large {\it
positive} $x$ is
\begin{eqnarray}
\eta &=& \arctan \frac{x}{b}+\eta _0 e^{2K(x)}\nonumber \\
&&+\frac{2m}{\hbar ^{2} k_r} \int_0^{x}\left[\epsilon - |\Delta
(x^\prime ) |\, \frac{b}{|x^\prime |}\right] e^{2K(x)-2K(x^\prime
)}\, dx^\prime \quad \label{eta-general}
\end{eqnarray}
where $\eta _0$ is a constant and
\[
K(x)=m\hbar ^{-2} k_r^{-1}\int _0^x |\Delta (x^\prime )|\,
dx^\prime \ .
\]
For $\eta _0\ll 1$ and $\epsilon \ll \Delta _0$, the function
$\eta$ is close to $\pi /2$ for $b\ll x\lesssim x_0$ where $x_0
\sim \xi \ln (1/|\gamma |)$ and
\begin{equation}
\gamma =2m\hbar ^{-2} k_r^{-1}\int_0^{\infty}\left[\epsilon -(b/x)
|\Delta|\right] e^{-2K(x )}\, dx \ . \label{gamma-def}
\end{equation}
$\gamma$ measures the distance from a CdGM level: $\gamma =0$
exactly when $\epsilon =\epsilon _{\mu}(k_z)$. The validity of
Eq.\ (\ref{eta-general}) is restricted by the condition $\gamma
\ll 1$ which generally holds if $\epsilon \ll \Delta _0$. The
function $\eta$ grows with $|x|$ at distances $x \gtrsim x_0$. To
find its behavior in the region $\eta \sim 1$ we can neglect small
terms with $\epsilon$ and $b/x$ in Eq.\ (\ref{eq-eta2}). The
solution is
\begin{equation}
\tan \left(\frac{\eta}{2} -\frac{\pi}{4}\right) = Ce^{2K(x)} \ .
\label{eta-nonlin}
\end{equation}
Matching with Eq.\ (\ref{eta-general}) at $\xi \ll x \ll x_0$
gives $C=\left(\gamma +\eta _0\right)/2$. For $\gamma +\eta _0 >
0$, the function $\eta \rightarrow 3\pi /2$ while $w$ diverges
exponentially as $x\rightarrow \infty$. If $\gamma +\eta _0 < 0$,
the function $\eta $ approaches $-\pi /2$, and $w$ diverges again.
However, if $\gamma +\eta _0 = 0$, the value $\eta =\pi /2$ is
stable (see Fig. \ref{fig-eta}) and the wave function decays for
$x\rightarrow \infty$.

The solution at $|x|\ll x_0$ for {\it negative} $x$ is obtained
from Eq.\ (\ref{eta-general}) by replacing $K(x)$ with $-K(x)$.
The function $\eta$ is close to $-\pi /2$ for $b\ll |x|\ll x_0$.
Its behavior for $|x| \gtrsim x_0$ is determined by Eq.\
(\ref{eta-nonlin}) where $C=2/\left(\gamma -\eta _0\right)$.
Equation (\ref{eta-nonlin}) then shows that $-3\pi /2< \eta
(-\infty)<-\pi /2$ if $\gamma -\eta _0>0$.  The function $\eta$
grows and approaches $\eta (-\infty) \approx -3\pi /2 $ while $w$
diverges.  Similarly, $-\pi /2< \eta (-\infty)<\pi /2$ if $\gamma
-\eta_0<0$ so that $w$ also diverges. The value $\eta =-\pi /2$ is
stable if only $\gamma -\eta _0=0$. The wave function thus decays
at both ends if $\gamma =\eta _0 =0$, which corresponds to a
standard CdGM state.

{\it Reflection and transmission probabilities.}--
For a superconducting slab with a thickness $d$ we thus have a
linear combination of two solutions $ \hat w =A_>\hat w^{>}+
A_<\hat w^{<} $ where $A_>$ and $A_<$ are constants. The first
solution $w_{1,2}^{>}$ has $\eta _0 =-\gamma$ and decays at
$x\rightarrow \infty$. Thus $\eta _> (+|x|)=\pi /2$ for $x\gg \xi$
while $\eta _> (-|x|)$ satisfies
\begin{equation}
\tan \left[\frac{1}{2} \eta _>(-|x|)-\frac{\pi}{4}\right] =\gamma
^{-1 }e^{-2K(|x|)} \label{eta>}
\end{equation}
according to Eq.\ (\ref{eta-nonlin}). For $\gamma \ne 0$ the phase
$\eta _> =-3\pi /2 +2\pi n$ at $x =-|x| \rightarrow -\infty$. The
amplitude factor for $x=-|x|$ is found from Eqs.\ (\ref{eq-zeta2})
and (\ref{eta-nonlin})
\[
\zeta _>(-|x|)=K(|x|)+\frac{1}{2}\ln\left(\frac{\gamma ^2 +
e^{-4K(|x|)}}{1+\gamma ^2}\right) \ .
\]
For $x=|x|$ it is simply  $\zeta _>(+|x|)= -K(|x|) $. The other
solution $w_{1,2}^< (x)= w_{1,2}^{>*} (-x)$ grows at $x\rightarrow
+\infty$.

The particle transmission $D_e$ and hole (Andreev) reflection
$R_h$ probabilities, $ 1=R_h + D_e$, are determined such that
$D_e=\left| w_1(z=d)/w_1(z=0)\right|^2$ provided there are no
transmitted holes, $w_2=0$ at $z=d$. We denote  the
$x$-coordinates of the end points of the trajectory at $z=0$ and
$z=d$ as $x_-$ and $x_+$, respectively, such that $d\tan \theta
=x_+-x_-$. For trajectories crossing the vortex axis, $0<x_+<d\tan
\theta$, $x_- =-|x_-|$, we find
\begin{equation}
D_e=\left( \gamma ^2 +a^2\right)^{-1}\,
\cosh^{-2}\left[K(|x_+|)+K(|x_-|)\right] \label{transmittion}
\end{equation}
where
\[
a=\cosh
\left[K(|x_+|)-K(|x_-|)\right]/\cosh\left[K(|x_+|)+K(|x_-|)\right]
\]
is the half-width of the level $\gamma =0$ proportional to the
escape rate of excitations from the level through the gapped
region far from the vortex. For $\gamma \rightarrow 0$, the
transmission coefficient becomes $D_p=\cosh
^{-2}\left[K(|x_+|)-K(|x_-|)\right] $. It is $D_p=1$ for resonant
trajectories that go through the middle of the vortex,
$|x_-|=|x_+|=(d/2)\tan \theta$. For trajectories that do not cross
the vortex axis, $x_-, x_+<0$ or $x_-
>0, x_+>d\tan \theta$, we find
$ D_p=\cosh ^{-2}\left[K(x_+)-K(x_-)\right] $ where $x_+ -x_-
=d\tan \theta$ .

Equation (\ref{transmittion}) suggests that the largest
contribution to the transmission through the energy gapped region
comes from the resonant tunneling when the energy  coincides
exactly with one of the levels in the vortex core. The
transmission probability is then unity if the trajectory crosses
the vortex close to the half of its length. However, the width of
the resonance $a$ is exponentially small which leads to a small
number of transmitted particles. This exponent, however,
corresponds to only a half of the slab thickness, not to the
entire slab thickness as it would be without vortices. The
exponential decay of the number of transmitted particles
disappears for trajectories which go almost parallel to the vortex
axis because the projection of the trajectory on the plane
perpendicular to the vortex axis shrinks as $\theta \rightarrow
0$. The transport is thus determined by trajectories almost
parallel to the vortex axis; its exponential dependence on the
slab thickness is replaced with a power-law behavior.

{\it The heat current.}--
The energy current along $z$ is
\begin{eqnarray*}
I_{{\cal E}} \!&=&\!\!\! \int  d^2r \sum _\mu \int \frac{d
k_z}{2\pi m}\, \left[\epsilon _\mu \, u^*_{\mu k_z}
\left(\hbar k_z-\frac{e}{c}A_z\right)u_{\mu k_z} n(\epsilon_\mu )\right. \\
&&\!\! -\left. \epsilon_\mu \, v^*_{\mu k_z} \left(\hbar k
_z+\frac{e}{c}A_z\right)v_{\mu k_z} \left[1-n(-\epsilon_\mu
)\right] \right] \ .
\end{eqnarray*}
Particles $u^*u$ with the distribution $n(\epsilon)$ carry the
energy $+ \epsilon$ while the holes $v^*v$ with the distribution
$1-n(-\epsilon)$ carry the energy $-\epsilon$. If the electrodes
are in equilibrium, $1-n(-\epsilon) =n(\epsilon)$ in each
electrode.

Consider first an infinitely thick slab. Using Eq. (\ref{ident1})
the energy current between the two electrodes becomes
\[
I_{{\cal E}}=\sum _\mu \int \epsilon _\mu n(\epsilon_\mu )\,
\frac{\partial \epsilon _\mu }{\partial k_z}\, \frac{d k_z}{2\pi
\hbar} \ .
\]
Excitations with positive group velocity $\hbar ^{-1}\partial
\epsilon _\mu /\partial k_z$ have the distribution
$n_1=[e^{\epsilon /T_1} +1]^{-1}$ as in the electrode 1. For those
with negative group velocity the distribution is $n_2=[e^{\epsilon
/T_2} +1]^{-1}$ as in the electrode 2. Therefore
\begin{equation}
I_{{\cal E}}=\sum _\mu \int_{p_z>0} \epsilon _\mu \left[
n_1(\epsilon_\mu )-n_2(\epsilon_\mu )\right]\, \left|
\frac{\partial \epsilon _\mu }{\partial k_z}\right|\, \frac{d
k_z}{2\pi \hbar} \ . \label{heatcurrent-Land}
\end{equation}
By the order of magnitude, the heat current is $I_{{\cal E}}\sim
(T/\hbar)(k_F\xi )(T/T_c)(T_1-T_2)$ with the thermal conductance
\begin{equation}
\kappa \sim (T/\hbar)(k_F\xi )(T/T_c) \label{heatcond-Land}
\end{equation}
in compliance with Eq. (\ref{Landauer-cond}).

For a finite slab thickness, $I_{\cal E}$ can be expressed through
the transmission and reflection coefficients,
\begin{eqnarray*}
I_{\cal E}\!\! &= &\!\!\sum \int d\epsilon |v_z|\left[\epsilon
n_1(\epsilon) - \epsilon
 R_h \left[ 1-n_1(-\epsilon)\right] -\epsilon  D_e
n_2(\epsilon)\right]\\
 \!&=&\! 2\nu _F \int _{v_z>0}
\frac{d\Omega}{4\pi} |v_z| \int dx_-\, db \int  D_e\, \epsilon
\left( n_1 -n_2\right)\, d\epsilon \ ,
\end{eqnarray*}
where the sum is over all the trajectories; $\nu_F$ is the
single-spin density of states at the Fermi level. The first two
terms in the upper line are due to incoming particles and Andreev
reflected holes on one side of the slab. The third term is due to
transmitted particles from the other side.

Using Eq.\ (\ref{transmittion}) we find that for the resonant
trajectories $-d\tan \theta <x_-<0$ the integration over $dx_-$
selects $|x_-|$ close to $|x_+|$. The angles along the vortex core
axis are small such that $k_r=k_F\sin \theta \ll k_F$ and
$K(x)=(m/2\hbar ^2 k_r)\Delta ^\prime (0) x^2$. The localization
radius of the wave function in the $x$ direction is $\lambda =
\hbar \left(k_F/\pi m\Delta ^\prime (0)\right)^{1/2}\theta
^{1/2}\sim \xi \theta ^{1/2}$. Let us put
$x_0=\left(|x_-|-|x_+|\right)/2 $ while $|x_-|+|x_+|=d\tan
\theta$. For small $\theta$ we have $K(|x_+|)-K(|x_-|)= x_0\, (md
/\hbar^2 k_F )\Delta ^\prime (0)$ and $K(|x_+|)+K(|x_-|)\sim
\theta \, (md^2 /\hbar^2 k_F)\Delta ^\prime (0)\sim d^2\theta /\xi
^2$. Therefore, $x_0\sim \xi ^2/d$ and $\theta \sim (\xi /d)^2$.
As a result,
\begin{equation}
I_{{\cal E}}\sim \nu _F\hbar ^4 v_F^5 d^{-6}\left(\Delta ^\prime
(0)\right)^{-4}\left(T^2_1-T^2_2\right) \ . \label{heatcurrent}
\end{equation}
Trajectories that do not cross the vortex but pass within an angle
$\theta \sim \xi ^2/d^2$ near its axis give the same order of
magnitude. The thermal conductance is $ \kappa \sim (T/\hbar
)(k_F\xi )^2(\xi /d)^6 $. The semi-classical approach requires
$k_r\lambda \gg 1$, which restricts the angle by $\theta \gg
(k_F\xi )^{-2/3}$. As a result, Eq.\ (\ref{heatcurrent}) applies
for $d/\xi \ll (k_F\xi)^{1/3}$. However, transition to the
Landauer equation (\ref{heatcond-Land}) occurs already at smaller
$d$. Thus the crossover region has be considered separately,
taking into account non-quasiclassical corrections to the wave
functions.

In conclusion, we have shown that with respect to the
single-electron transport the vortex core in a thick clean
superconductor is equivalent to a mesoscopic conducting channel
where the conductance is determined by the Landauer formula with
CdGM states playing the role of transverse modes. For a finite
slab thickness $d$, the vortex core behaves like an Andreev wire:
the thermal conductivity along it drops off as $d^{-6}$ due to a
finite order parameter everywhere except at the vortex axis.


This work was supported in part by the US DOE Office of Science
under contract No. W-31-109-ENG-38, by Russian Foundation for
Basic Research grant 02-02-16218, the Program ``Quantum
Macrophysics'' of the Russian Academy of Sciences, and NATO
Collaborative Linkage Grant No. PST.CLG.978122. A.S.M. is grateful
to the Low Temperature Laboratory at the Helsinki University of
Technology for hospitality.



\end{document}